\documentclass[12pt]{article}%
\usepackage{amsmath,latexsym}
\usepackage{graphicx}
\usepackage{amsmath}
\usepackage{amsfonts}
\usepackage{amssymb}%
\begin{document}

\title{{Revisiting wormholes supported by two
   non-interacting fluids}}
   \author{
Peter K.F. Kuhfittig*\\  \footnote{kuhfitti@msoe.edu}
 \small Department of Mathematics, Milwaukee School of
Engineering,\\
\small Milwaukee, Wisconsin 53202-3109, USA}

\date{}
 \maketitle

\begin{abstract}\noindent
This paper extends several previous studies
of wormholes supported by two non-interacting
fluids beginning with a combined model of ordinary
matter and phantom dark energy with an anisotropic
matter distribution.  After noting that such
wormholes could only exist on very large scales,
the two-fluid model is extended to neutron stars,
previously treated only for the isotropic case.
The model is completed by incorporating certain
generic features proposed by the author in an
earlier study.  

\end{abstract}

\section{Introduction}\label{E:introduction}

Wormholes are handles or tunnels in spacetime
connecting widely separated regions of our
Universe or even entirely different universes
in a multiverse.  Apart from some forerunners,
macroscopic traversable wormholes were first
studied in detail by  Morris and Thorne
\cite{MT88} in 1988.  They had proposed the
following static and spherically symmetric
line element for a wormhole spacetime:
\begin{equation}\label{E:line}
ds^{2}=-e^{2\Phi(r)}dt^{2}+e^{2\Lambda(r)}dr^2
+r^{2}(d\theta^{2}+\text{sin}^{2}\theta\,
d\phi^{2}),
\end{equation}
where $e^{2\Lambda(r)}=1-b(r)/r$.  (We are
using units in which $c=G=1$.)  In the
terminology introduced in Ref. \cite{MT88},
$\Phi=\Phi(r)$ is called the \emph{redshift
function}, which must be everywhere finite
to prevent an event horizon.  The function
$b=b(r)$ is called the \emph{shape function}
since it determines the spatial shape of the
wormhole when viewed, for example, in an
embedding diagram \cite{MT88}.  The spherical
surface $r=r_0$ is called the \emph{throat}
of the wormhole, where $b(r_0)=r_0$.  The
shape function must also meet the requirement
$b'(r_0)<1$, called the \emph{flare-out
condition}, while $b(r)<r$ for $r>r_0$.  We
also require that $b'(r_0)>0$.  In classical
general relativity, the flare-out condition
can only be met by violating the null energy
condition (NEC), which states that for the
energy-momentum tensor $T_{\alpha\beta}$,
\begin{equation}
  T_{\alpha\beta}k^{\alpha}k^{\beta}\ge 0
  \,\,\text{for all null vectors}\,\,
  k^{\alpha}.
\end{equation}
In particular, for the outgoing null vector
$(1,1,0,0)$, the violation becomes
\begin{equation}
   T_{\alpha\beta}k^{\alpha}k^{\beta}=
   \rho +p_r<0.
\end{equation}
Here $T^t_{\phantom{tt}t}=-\rho$ is the energy
density, $T^r_{\phantom{rr}r}= p_r$ is the
radial pressure, and
$T^\theta_{\phantom{\theta\theta}\theta}=
T^\phi_{\phantom{\phi\phi}\phi}=p_t$ is
the lateral (transverse) pressure.

Matter that violates the NEC is called
``exotic" in Ref. \cite{MT88}.  So in
classical general relativity, the
violation of the NEC is equivalent to
the need for exotic matter, at least in
the vicinity of the throat.  It follows
that exotic matter can only be avoided
by departing from a strict classical
setting.  Examples are an appeal to
$f(R)$ modified gravity \cite{LO09} or
to the existence of an extra spatial
dimension \cite{pK18}.

The idea that a wormhole can be supported
by two non-interacting fluids is not new
\cite{KRG10, pK13, RRI13}.  Using the
superscripts 1 and 2 to represent the two
fluids, the energy-momentum tensor takes
on the form
\begin{equation}
   T^t_{\phantom{tt}t}=-(\rho^1+\rho^2),
\end{equation}
\begin{equation}
    T^r_{\phantom{tt}r}=p_r^1+p_r^2,
\end{equation}
and
\begin{equation}
   T^\theta_{\phantom{\theta\theta}\theta}=
   T^\phi_{\phantom{\phi\phi}\phi}=
   p_t^1+p_t^2.
\end{equation}
Referring now to line element (\ref{E:line}),
the Einstein field equations are
\begin{equation}\label{E:Einstein1}
  \frac{b'}{r^2}=8\pi\,(\rho^1+\rho^2),
\end{equation}
\begin{equation}\label{E:Einstein2}
   -\frac{b}{r^3}+
   2\left(1-\frac{b}{r}\right)\frac{\Phi'}{r}
   =8\pi\,(p_r^1+p_r^2),
\end{equation}
and
\begin{equation}\label{E:Einstein3}
   \left(1-\frac{b}{r}\right)
   \left[\Phi''-\frac{b'r-b}{2r(r-b)}\Phi'
   +(\Phi')^2+\frac{\Phi'}{r}-
   \frac{b'r-b}{2r^2(r-b)}\right]
   =8\pi(p_t^1+p_t^2).
\end{equation}

Ref. \cite{KRG10} discusses quintom
wormholes, so that the two fluids are
quintessence and phantom dark energy.  In
Ref. \cite{pK13}, dealing with neutron
stars, the matter sources are neutron and
quark matter, respectively.  In the third
case, to be discussed further below, the
two non-interacting fluids are ordinary
matter and phantom dark energy \cite{RRI13}.

The main purpose of this paper is to
compare and contrast the last two cases.

\section{Background}\label{S:Background}

As noted in the Introduction, we are going to
be primarily interested in the models discussed
in Refs. \cite{pK13} and \cite{RRI13}.  Since
these describe very different physical
situations, some preliminary remarks are going
to be needed.

Our first concern is the extreme radial tension
at the throat of a Morris-Thorne wormhole,
discussed in Ref. \cite{MT88}.  We first  need
to recall that the radial tension $\tau(r)$ is
the negative of the radial pressure $p_r(r)$.
According to Ref. \cite{MT88}, the Einstein
field equations can be rearranged to yield
\begin{equation}\label{E:tau}
   \tau(r)=\frac{b(r)/r-2[r-b(r)]\Phi'(r)}
   {8\pi Gc^{-4}r^2},
\end{equation}
temporarily reintroducing $c$ and $G$.  So
the radial tension at the throat becomes
\begin{equation}\label{E:tension}
   \tau(r_0)=\frac{1}{8\pi Gc^{-4}r_0^2}\approx
   5\times 10^{41}\frac{\text{dyn}}{\text{cm}^2}
   \left(\frac{10\,\text{m}}{r_0}\right)^2.
\end{equation}
It is noted in Ref. \cite{MT88} that  for
$r_0=3\,\,\text{km}$, $\tau$ has the same
magnitude as the pressure at the center of
a massive neutron star.  It follows from
Eq. (\ref{E:tension}) that the widely
discussed wormhole solutions based on dark
matter and dark energy could only exist on
very large scales, i.e., by possessing
very large throat radii $r_0$.  For further
discussion of this problem, see Ref.
\cite{pKsurvey}.

Before continuing, we need to consider two
other topics discussed in Ref. \cite{pK20}.
The first is the shape function
\begin{equation}\label{E:shape}
   b(r)=r_0\left(\frac{r}{r_0}
      \right)^{\alpha},\quad 0<\alpha<1,
\end{equation}
introduced in Ref. \cite{MT88} and used
in Ref. \cite{RRI13}.  Ref. \cite{pK20}
defines a shape function to be \emph{typical}
if it meets the following requirements:
$b(r_0)=r_0$, $0<b'(r_0)<1$, and $b(r)$ is
concave down near the throat.  It is shown
that any such shape function can be
approximated by Eq. (\ref{E:shape}).  So
this form of the shape function is
essentially generic.

Having defined a typical shape function,
Ref. \cite{pK20} goes on to discuss the
nature of exotic matter in a Morris-Thorne
wormhole.  First we need to recall that
phantom dark energy is characterized by
the equation of state $p=\omega\rho$,
$\omega <-1$, and can therefore support
traversable wormholes \cite{oZ05}.  The
use of Eq. (\ref{E:shape}) leads to the
surprising conclusion that the converse
is also true in the following sense: if
we are dealing with a typical shape
function, the equation of state of exotic
matter in the vicinity of the throat is
given by $p_r=\omega\rho$, $\omega <-1$,
where $p_r$ is the radial pressure.

In this paper, it turns out to be
convenient to use the form
\begin{equation}\label{E:EoS1}
   p_r=-\omega\rho, \quad \omega >1.
\end{equation}

\section{The combined model}
   \label{S:combined}

In this section, we will examine the
combined model in Ref. \cite{RRI13}
comprising ordinary matter and phantom
dark energy.  Here ordinary matter has
the standard perfect-fluid equation of
state
\begin{equation}\label{E:perfect}
   p=m\rho, \quad 0<m<1.
\end{equation}
The phantom-energy density $\rho^{ph}$
is assumed to have the form
\begin{equation}\label{E:ph}
   \rho^{ph}=n\rho, \quad n>0.
\end{equation}
This is a natural assumption but subject
to the condition that $n$ be extremely
small.  For the phantom-energy case, we
have from Eq. (\ref{E:EoS1}) that
\begin{equation}\label{E:EoS2}
   p_r^{ph}=-\omega\rho^{ph}, \quad
      \omega >1.
\end{equation}
So in Eqs.
(\ref{E:Einstein1})-(\ref{E:Einstein3}),
the superscript 1 stands for ordinary
matter and the superscript 2 for phantom
energy, i.e., $\rho^1=\rho$ and $\rho^{ph}
=n\rho$.  Substituting Eq. (\ref{E:shape})
in Eq. (\ref{E:Einstein1}) yields
\begin{equation}\label{E:rho}
   \rho=\frac{\alpha}{8\pi(1+n)r_0^2}
   \left(\frac{r}{r_0}\right)^{\alpha -3},
\end{equation}
which leads at once to
\begin{equation}
    p=m\rho=\frac{m\alpha}{8\pi(1+n)r_0^2}
    \left(\frac{r}{r_0}\right)^{\alpha -3}.
\end{equation}
and
\begin{equation}
   p_r^{ph}=-\omega\rho^{ph}=-\omega n\rho=
   -\frac{\omega n\alpha}{8\pi(1+n)r_0^2}
   \left(\frac{r}{r_0}\right)^{\alpha -3}.
\end{equation}
The expression for $p_t^{ph}$ is given
in Ref. \cite{RRI13}.  Since $p_t^{ph}
\neq p_r^{ph}$, the pressure is
anisotropic.

Since we are dealing with a combined
model, we need to introduce the following
notations: the effective density is
denoted by $\rho^{eff}=\rho+\rho^{ph}$
and the effective radial pressure by
$p_r^{eff}=p_r+p_r^{ph}$.  So
\begin{equation}\label{E:rhoeff}
   \rho^{eff}=\frac{\alpha}{8\pi(1+n)r_0^2}
   \left(\frac{r}{r_0}\right)^{\alpha -3}
   +\frac{n\alpha}{8\pi(1+n)r_0^2}
   \left(\frac{r}{r_0}\right)^{\alpha -3}
   =\frac{\alpha}{8\pi r_0^2}
   \left(\frac{r}{r_0}\right)^{\alpha-3}.
\end{equation}
Since Eq. (\ref{E:perfect}) represents a
perfect fluid, we have $p_r=m\rho$ and hence
\begin{multline}\label{E:peff}
   p_r^{eff}=\frac{m\alpha}{8\pi(1+n)r_0^2}
   \left(\frac{r}{r_0}\right)^{\alpha -3}
   -\omega\frac{n\alpha}{8\pi(1+n)r_0^2}
   \left(\frac{r}{r_0}\right)^{\alpha-3}\\
   =\frac{\alpha(m-n\omega)}{8\pi(1+n)r_0^2}
   \left(\frac{r}{r_0}\right)^{\alpha-3}.
\end{multline}

We saw in the Introduction that for a
Morris-Thorne wormhole, meeting the
flare-out condition at the throat implies
that $\rho+p_r<0$.  Since we are dealing
with a more complicated model, we need to
show explicitly that $\rho^{eff}+p_r^{eff}
<0$.  So from Eqs. (\ref{E:rhoeff}) and
(\ref{E:peff}),
\begin{multline}\label{E:NEC}
   \left.\rho^{eff}+p_r^{eff}\right|_{r=r_0}
   =\left.\frac{1}{8\pi r^2}\left[\alpha +
   \frac{\alpha(m-n\omega)}{n+1}\right]
   \left(\frac{r}{r_0}\right)^{\alpha -3}
   \right|_{r=r_0}\\=\frac{\alpha(n+1+m-n\omega)}
   {n+1}<0
\end{multline}
provided that
\begin{equation}\label{E:omega1}
   \omega>\frac{1+n+m}{n}.
\end{equation}

Given that $n$ is extremely small,
the phantom-energy parameter $\omega$
is extremely large, which is completely
unphysical unless $r_0$ is also
extremely large:  in the latter case,
we can see from Eq. (\ref{E:rho}) that
Eq. (\ref{E:ph}) could hold for larger
$n$, say $n\ge 1$.

\emph{Remark:} That wormholes supported
by phantom energy could only exist on
large scales has already been noted
after Eq. (\ref{E:tension}).

\section{Neutron stars}\label{S:neutron}
We have seen that the two-fluid model
could only work for much larger $n$
and may therefore be applicable to
neutron stars.  Here we need to recall
that while neutrons are normally held
together by the strong force, the extreme
conditions near the center are likely to
cause the neutrons to become deconfined,
resulting in quark matter.  So in the
vicinity of the throat $r=r_0$, deep
in the interior, we can assume that both
neutron matter and quark matter are
present, thereby preserving the two-fluid
model.  (We will denote the density of
neutron matter by $\rho$ and the density
of quark matter by $\rho^{q}$.)

While the interior $r<r_0$ is not part
of the wormhole spacetime, the highly
dense quark matter near the center still
contributes to the gravitational field.
This can be compared to a thin-shell
wormhole from a Schwarzschild black hole
\cite{PV95}.  Here the black hole
generates the gravitational field.
Using the MIT bag model, it is shown in
Ref. \cite{pK13} that $b=b(r)$ qualifies
as a typical shape function due to the
presence of quark matter.  So we can
retain the shape function in Eq.
(\ref{E:shape}).

Our two-fluid model therefore
comprises neutron matter and quark
matter, assumed to be essentially
non-interacting and possessing an
anisotropic matter distribution in
order to be consistent with our
earlier discussion.  This assumption
leads directly to the analogue of
Eq. (\ref{E:EoS2}):
\begin{equation}
   p_r^{q}=-\omega\rho^{q}, \quad
      \omega >1,
\end{equation}
where the superscript $q$ refers to
quark matter.  We also need to recall
from the end of Sec. \ref{S:Background}
that the phantom-energy equation of
state has a converse:  matter that
violates the NEC can be assumed
to have the form in Eq. (\ref{E:EoS1}).

The rest of this paper is devoted to
showing that these assumptions are
reasonable.  To that end, we first note
that the density of neutron matter ranges
from $3.7\times 10^{17}\,\text{kg}/\text{m}^3$
to $5.9\times 10^{17}\,\text{kg}/\text{m}^3$
and that the density of quark matter is
approximately
$2.7\times 10^{18}\,\text{kg}/\text{m}^3$.
Denoting the former by $\rho$ and the
latter by $\rho^{q}$, the analogue of
Eq. (\ref{E:ph}) becomes
\begin{equation}
   \rho^{q}=n\rho, \quad n\approx 6.
\end{equation}
Eq. (\ref{E:omega1}) now reads
\begin{equation}\label{E:omega2}
   \omega>\frac{1+6+m}{6},
\end{equation}
which is acceptable from a physical
standpoint.  Furthermore, from
Eq. (\ref{E:peff}), we find that at
$r=r_0$ with $n=6$
\begin{equation}
   p_r^{eff}=\frac{\alpha(m-6\omega)}{7}
   \frac{1}{8\pi r_0^2}\frac{c^4}{G}=
   \frac{\alpha(m-7-m)}{7}\frac{1}{8\pi r_0^2}
   \frac{c^4}{G}
\end{equation}
from Eq. (\ref{E:omega1}).  So
\begin{equation}
   p_r^{eff}=-\alpha\frac{1}{8\pi r_0^2}
   \frac{c^4}{G}=\alpha\tau(r_0).
\end{equation}
To be consistent with Eq. (\ref{E:tension}),
our generic shape function (\ref{E:shape})
must be restricted in an obvious way:
$\alpha$ must be closer to unity, say
$\alpha>0.5$.

That neutron stars can support
traversable wormholes had already been
proposed in Ref. \cite{pK13} under the
assumption of isotropic pressure.
Neutron stars connected by wormholes
are also discussed in Ref. \cite{vD12}.

The above discussion suggests that
wormholes with moderate throat sizes
are actually compact stellar objects,
made explicit in Ref. \cite{pK22}.

The results in this section depend
on Eq. (\ref{E:omega2}), an assumption
that has proved to be physically
reasonable.  Further confirmation
comes from the gradient of the
redshift function, discussed in
the next section.

\section{The gradient of the redshift
   function}\label{E:red}

Combining Eqs. (\ref{E:Einstein1}) and
(\ref{E:Einstein2}), we obtain
\begin{multline}
   \frac{rb'-b}{r^3}+2\left(1-\frac{b}{r}\right)
   \frac{\Phi'}{r}=8\pi(\rho+\rho^{q}+p_r+p_r^{q})
   =8\pi(\rho+n\rho+m\rho-\omega n\rho)\\
   =8\pi\frac{p}{m}(1+n+m-\omega n)
\end{multline}
since $p=m\rho$.  Solving for $\Phi'$, we get
\begin{equation}\label{E:gradient}
   \Phi'=
   \frac{1}{r[r-b(r)]}\left[-\frac{rb'(r)-b(r)}{2}
   +4\pi\frac{p}{m}(1+n+m-\omega n)r^3\right].
\end{equation}

As another check on the plausibility of our
model, let us return to line element
(\ref{E:line}): if $e^{2\Lambda(r)}=1-2m(r)/r$
with $m(0)=0$, the line element represents a
stellar model \cite{MTW}.  Here the parameter
$\Phi$ is usually viewed as the relativistic
version of the Newtonian potential $\Phi=
-M/r$ \cite{MTW}:
\begin{equation}
   \frac{d\Phi}{dr}=\frac{M+4\pi r^3p}
      {r(r-2M)},
\end{equation}
where $M$ is the mass of the star.  So
$d\Phi/dr=M/r^2$ in the Newtonian limit.  If
we let $b(r)=2M$ in Eq. (\ref{E:gradient}),
we get
\begin{equation}
   \Phi'=\frac{1}{r(r-2M)}\left[M+
   4\pi\frac{p}{m}(1+n+m-\omega n)r^3\right].
\end{equation}
To complete the comparison, we must also have
\begin{equation}
    \frac{1+n+m-\omega n}{m}=1,
\end{equation}
which implies that $\omega=1+1/n$; thus
$\omega\approx 7/6$, since $n\approx 6$.
So $\omega$ is only slightly larger than
unity.

\section{Conclusion}
Wormholes supported by two non-interacting
fluids are discussed in Refs. \cite{KRG10,
pK13, RRI13}.  This paper begins with the
combined model of ordinary matter and
phantom dark energy with an anisotropic
matter distribution, discussed in Ref.
\cite{RRI13}.  While leading to a valid
wormhole solution, such wormholes could
only exist on very large scales, i.e.,
with very large throat sizes, shown
previously in Ref. \cite{pKsurvey}.

Extending this model to neutron stars
calls for two non-interacting (or at
least weakly interacting) fluids,
neutron matter and quark matter.  Both
are assumed to be present near the
throat, being close enough to the
center of the neutron star.   The
two-fluid model carries over directly
to neutron stars provided that the equation
of state $p_r^{ph}=-\omega\rho^{ph}$,
$\omega >1$, for phantom dark energy can be
replaced by the equation of state
$p_r^{q}=-\omega\rho^{q}$, $\omega>1$,
for quark matter.  This replacement can
be justified as follows: it was noted in
Sec. \ref{S:Background} that our shape
function can be approximated by
$b(r)=r_0(r/r_0)^{\alpha}$,
$0<\alpha<1$ \cite{pK20}, thereby meeting
the flare-out condition.  Since we are
going beyond a basic Morris-Thorne
wormhole, the violation of the NEC
needs to be checked separately; this
is verified in Eqs. (\ref{E:NEC}) and
(\ref{E:omega1}).  Moreover, it is
shown in Ref. \cite{pK20} that matter
that violates the NEC can be assumed to
have the form $p_r=-\omega\rho$,
$\omega>1$, enough to suggest that the
analogous equation of state for quark
matter is $p_r^q=-\omega\rho^q$,
$\omega>1$.  It is shown in Sec.
\ref{S:neutron} that this is indeed
is a reasonable assumption, thereby
preserving the two-fluid model.  It is
interesting to note that the resulting
radial pressure turns out to be
$p_r^{eff}=\alpha\tau(r_0)$.  To be
consistent with Eq. (\ref{E:tension}),
$\alpha$ must be sufficiently large.

Returning now to the parameter $\omega$,
we saw in Sec. \ref{S:combined} that
the combined model comprising ordinary
matter and phantom dark energy leads to
a large physically unacceptable value
of the parameter $\omega$.  By contrast,
for neutron stars, an examination of the
gradient of the redshift function,
discussed in Sec. \ref{E:red}, shows that
$\omega$ is only slightly larger than unity.\

The result is a viable extension of the
two-fluid model in Ref. \cite{RRI13} to
neutron stars.  This outcome suggests that
wormholes with moderately-sized throat
sizes are actually compact stellar objects,
made explicit in Ref. \cite{pK22}.

Neutron stars connected by wormholes are also
discussed in Ref. \cite{pK13}, assuming
an isotropic matter distribution.  It is
noted in Ref. \cite{vD12} that an observable
tell-tale sign of such a wormhole would be
any observed variation in the mass of the
neutron star.  The existence of compact
stellar objects with an anisotropic matter
distribution is also discussed in Refs.
\cite{zY18, zY22} in $f(R, T)$ modified
gravity.  The detection of phantom-energy
wormholes, as well as phantom-energy black
holes, by means of gravitational lensing
is discussed in Ref. \cite{aO19}.


\begin{thebibliography}{20}
\bibitem{MT88}M. S. Morris and K. S. Thorne,
   Wormholes in spacetime and their use for
   interstellar travel: A tool for teaching
   general relativity, American Journal of Physics
   56 (1988), 395-412.
\bibitem{LO09}F. S. N. Lobo and M. A. Oliveira,
   Wormhole geometries in $f(R)$ modified theories
   of gravity, Physical Review D 80 (2009),
   ID: 104012.
\bibitem{pK18}P. K .F. Kuhfittig, Traversable
   wormholes sustained by an extra spatial
   dimension, Physical Review D 98 (2018),
   ID: 064041.
\bibitem{KRG10}P. K. F. Kuhfittig, F. Rahaman
   and A. Ghosh, Quintom wormholes, International
   Journal of Theoretical Physics 49 (2010),
   1222-1231.
\bibitem{pK13}P. K. F. Kuhfittig, Neutron star
   interiors and topology change, Advances in
   Mathematical Physics 2013 (2013), ID: 630196.
\bibitem{RRI13}F. Rahaman, S. Ray and S. Islam,
   Wormholes supported by two non-interacting
   fluis, Astrophysics and Space Science 346
   (2013), 245-252.
\bibitem{pKsurvey}P. K. F. Kuhfittig, A survey
   of recent studies concerning the extreme
   properties of Morris-Thorne wormholes,
   arXiv: 2202.07431 [gr-qc].
\bibitem{pK20}P. K. F. Kuhfittig, On the nature
   of exotic matter in Morris-Thorne wormholes,
   New Horizons in Mathematical Physics 4 (2020),
   29-32.
\bibitem{oZ05}O. B. Zaslavskii, Exacly solvable
   model of a wormhole supported by phantom
   energy, Physical Review D 72 (2005), ID: 061303.
\bibitem{PV95}E. Poisson and M. Visser, Thin-shell
   wormholes: Linearized stability, Physical
   Review D 52 (1995), ID: 7318.
\bibitem{vD12}V. Dzhunushaliev, V. Folomeev,
   B. Kleihaus and J. Kunz, Mixed
   neutron-star-plus-wormhole systems: Equilibrium
   configurations, Physical Review D 85 (2012),
   ID: 124028.
\bibitem{pK22}P. K. F. Kuhfittig, A note on
   wormholes as compact stellar objects, Fundamental
   Journal of Modern Physics 17 (2022), 63-70.
\bibitem{MTW}C. W. Misner, K. S. Thorne and
   J. A. Wheeler, Gravitation (New York:
   W. H. Freeman, 1973), chapter 23.
\bibitem{zY18}Z. Yousaf, M. Zaeem-ul-Haq
   Bhatti and M. Ilyas, Existence of compact
   structures in $f(R,T)$ gravity, European
   Physical Journal C 78 (2018), ID: 307.
\bibitem{zY22}Z. Yousaf, Kazuharu Bamba, M. Z.
   Bhatti and U. Farwa, Quasi static evolution
   of compact objects in modified gravity,
   General Relativity and Gravitation 54 (2022),
   ID: 7.
\bibitem{aO19}A. Ovgun, G. Gyulchev and
   K. Jusufi, Weak gravitational lensing by
   phantom black holes and phantom wormholes
   using the Gauss-Bonnet theorem, Annals of
   Physics 406 (2019), 152-172.


\end{thebibliography}
\end{document}